\documentclass[12pt]{article}
\usepackage{amssymb}
\usepackage{amsfonts}
\usepackage{amsmath}
\usepackage{epsfig}
\usepackage{graphicx}

\begin{document}

\title{Current algebra, statistical mechanics and quantum models}
\author{R. Vilela Mendes\thanks{%
rvilela.mendes@gmail.com; rvmendes@fc.ul.pt} \\
{\small CMAFCIO, Universidade de Lisboa,}\\
{\small Faculdade de Ci\^{e}ncias C6, 1749-016 Lisboa, Portugal}}
\date{ }
\maketitle

\begin{abstract}
Results obtained in the past for free boson systems at zero and nonzero
temperature are revisited to clarify the physical meaning of current algebra
reducible functionals which are associated to systems with density
fluctuations, leading to observable effects on phase transitions.

To use current algebra as a tool for the formulation of quantum statistical
mechanics amounts to the construction of unitary representations of
diffeomorphism groups. Two mathematical equivalent procedures exist for this
purpose. One searches for quasi-invariant measures on configuration spaces,
the other for a cyclic vector in Hilbert space. Here, one argues that the
second approach is closer to the physical intuition when modelling complex
systems. An example of application of the current algebra methodology to the
pairing phenomenon in two-dimensional fermion systems is discussed.
\end{abstract}

\section{Introduction}

All representations of canonical fields with a finite number of degrees of
freedom are equivalent to the Fock representation. However, for an infinite
number of degrees of freedom there are, in addition to the Fock
representation, infinitely many inequivalent representations of the
canonical commutation relations. In relativistic quantum field theory,
Haag's theorem states that, with a space-invariant vacuum, any
representation equivalent to Fock can only describe a free system.
Therefore, to obtain a non-trivial theory, one either works with a non-Fock
representation or with a Fock representation in a finite volume. In the
latter case one considers $N$ particles in a finite volume $V$. Calculations
are then carried out in the Fock representation, but in the end one may take 
$N,V\rightarrow \infty $ with the $N/V=\rho $ ratio fixed. The $N/V$ limit
thus provides a way to deal with non-trivial infinite systems using the Fock
representation. However, by the very nature of the fixed $\rho $ density
limit, it is unable to deal with systems with density fluctuations. This
shortcoming might be solved by the use of the reducible functionals to be
described later on.

In the field theory description of matter, the field operators $\psi \left(
x\right) $ and $\psi ^{\dag }\left( x\right) $ do not represent actual
physical observables. This, together with the strong uniqueness results on
the representation of the (finite-dimensional) canonical commutation
relations, were the original motivations for the proposal by Dashen, Sharp,
Callan and Sugawara \cite{Dashen} \cite{Sharp} \cite{Callan} \cite{Sugawara}
to use local density and current operators as descriptors of quantum
observables. Despite some early successes, mostly in the derivation of sum
rules, relativistic current algebra in space-time dimensions higher than $%
1+1 $ faced serious difficulties related to the non-finiteness of Schwinger
terms. By contrast, no such problem occurs for non-relativistic current
algebras which, already at a very early stage, have been proposed as a tool
for statistical mechanics \cite{Arefeva} \cite{Goldin1}. Nonrelativistic
current algebra was then extensively studied by Goldin and collaborators 
\cite{Goldin2} \cite{Goldin3} \cite{Goldin4}. A relation with the classical
Bogolubov generating functional has also been established, in particular as
a tool for constructing the irreducible current algebra representations \cite%
{Bogolubov1} \cite{Bogolubov2}. From a mathematical point of view, the early
considerations related to the $N/V$ limit have found a rigorous
interpretation in the framework of the infinite-dimensional Poisson analysis
in configuration spaces (\cite{KondratRMP} and references therein).

In this paper results obtained in the past for free boson systems at zero
and nonzero temperature are revisited with a view to clarify the physical
meaning and potential usefulness of current algebra reducible functionals.
Reducible functionals are associated to systems with density fluctuations,
which may lead to observable effects on phase transitions.

Using current algebra as a tool for the formulation of quantum statistical
mechanics is closely related to the problem of construction of unitary
representations of diffeomorphism groups. Two mathematical equivalent
procedures exist for this purpose. One searches for quasi-invariant measures
on configuration spaces, the other for a cyclic vector in Hilbert space.
Here, one argues that the second approach is closer to the physical
intuition when modelling complex systems. An example of application of the
current algebra methodology to the pairing phenomenon in two-dimensional
fermion systems is included.

\section{Boson gas, the infinite-dimensional Poisson measure and reducible
functionals}

\subsection{Infinite-dimensional Poisson measures and free Boson gases}

The framework of non-relativistic current algebra of many-body systems is a
particularly convenient way to establish the connection of the Boson gas
functional with infinite-dimensional measures, as well as to explore
generalizations. The basic variables of the many-body system are the smeared
currents \cite{Goldin1} \cite{Goldin2} (see also \cite{Goldin3} \cite%
{Goldin4} and references therein)%
\begin{eqnarray}
\varrho \left( f\right) &=&\int d^{3}xf\left( x\right) \varrho \left(
x\right)  \notag \\
\mathbf{J}\left( \mathbf{g}\right) &=&\int d^{3}x\mathbf{J}\left( x\right)
\bullet \mathbf{g}\left( x\right)  \label{2.1}
\end{eqnarray}%
$f\left( x\right) $ and $\mathbf{g}\left( x\right) $ being respectively
smooth compactly supported functions and smooth vector fields. The smeared
currents satisfy the infinite-dimensional Lie algebra,%
\begin{eqnarray}
\left[ \varrho \left( f\right) ,\varrho \left( h\right) \right] &=&0  \notag
\\
\left[ \varrho \left( f\right) ,\mathbf{J}\left( \mathbf{g}\right) \right]
&=&i\varrho \left( \mathbf{g\bullet \nabla }f\right)  \notag \\
\left[ \mathbf{J}\left( \mathbf{g}\right) ,\mathbf{J}\left( \mathbf{k}%
\right) \right] &=&i\mathbf{J}\left( \mathbf{k\bullet \nabla g-g\bullet
\nabla k}\right)  \label{2.2}
\end{eqnarray}%
each particular physical system corresponding to a different Hilbert space
representation of this algebra or of the semidirect product group generated
by the exponentiated currents%
\begin{eqnarray}
U\left( f\right) &=&e^{i\varrho \left( f\right) }  \notag \\
V\left( \phi _{t}^{\mathbf{g}}\right) &=&e^{it\mathbf{J}\left( \mathbf{g}%
\right) }  \label{2.3}
\end{eqnarray}%
$\phi _{t}^{\mathbf{g}}$ being the flow of the vector field $\mathbf{g}$%
\begin{equation}
\frac{d}{dt}\phi _{t}^{\mathbf{g}}\left( x\right) =\mathbf{g}\left( \phi
_{t}^{\mathbf{g}}\left( x\right) \right)  \label{2.4}
\end{equation}%
For a system of $N$ free bosons in a box of volume $V$, the normalized
ground state is%
\begin{equation}
\Omega _{N,V}\left( x_{1},\cdots ,x_{N}\right) =\left( \frac{1}{\sqrt{V}}%
\right) ^{N}  \label{2.5}
\end{equation}%
and the ground state functional is%
\begin{eqnarray}
L_{N,V}\left( f\right) &=&\left( \Omega _{N,V},U_{N,V}\left( f\right) \Omega
_{N,V}\right)  \notag \\
&=&\left( \frac{1}{V}\int_{V}d^{3}xe^{if\left( x\right) }\right) ^{N}
\label{2.6}
\end{eqnarray}%
Coupled with an equation of continuity relating $\varrho $ and $\mathbf{J}$,
this functional determines not only the representation of $U\left( f\right) $
but also that of $V\left( \phi _{t}^{\mathbf{g}}\right) $, up to a complex
phase multiplier that satisfies a cocycle condition\footnote{%
See Eq.(\ref{R30})}.

In the $N\rightarrow \infty $ limit with constant average density $\rho =%
\frac{N}{V}$ (also called the $N/V$ limit) one obtains%
\begin{eqnarray}
L\left( f\right) &=&\lim_{N\rightarrow \infty }\left( 1+\frac{\rho }{N}\int
\left( e^{if\left( x\right) }-1\right) d^{3}x\right) ^{N}  \notag \\
&=&\exp \left( \rho \int \left( e^{if\left( x\right) }-1\right) d^{3}x\right)
\label{2.7}
\end{eqnarray}%
which one recognizes as the characteristic functional of the
infinite-dimensional Poisson measure (see the Appendix).

Likewise the functional%
\begin{equation*}
L_{N/V}\left( f,\mathbf{g}\right) =\left( \Omega _{N/V},e^{i\varrho \left(
f\right) }e^{i\mathbf{J}\left( \mathbf{g}\right) }\Omega _{N/V}\right)
\end{equation*}%
is \cite{Menik1} in the $N/V$ limit%
\begin{equation*}
L\left( f,\mathbf{g}\right) =\exp \left( \rho \int \left\{ e^{if\left(
x\right) }\left( \det \partial _{m}\phi _{n}^{\mathbf{g}}\left( x\right)
\right) ^{1/2}-1\right\} d^{3}x\right)
\end{equation*}%
where $\det \partial _{m}\phi _{n}^{\mathbf{g}}\left( x\right) $ stands for
the Jacobian of the transformation $x\rightarrow \phi ^{\mathbf{g}}\left(
x\right) $.

Identifying $\rho d^{3}x$ in (\ref{2.7}) with the measure $d\mu $ in the
configuration spaces discussed in the Appendix, the $L\left( f\right) $
functional may also be written as a vacuum expectation functional. Expanding
the exponential in (\ref{2.7})%
\begin{equation}
L\left( f\right) =\sum_{n=0}^{\infty }\frac{e^{-\int d\mu }}{n!}\left( \int
e^{if\left( x\right) }d\mu \right) ^{n}  \label{2.8}
\end{equation}%
one may write%
\begin{equation}
L\left( f\right) =\left( \Omega ,U\left( f\right) \Omega \right)  \label{2.9}
\end{equation}%
for%
\begin{equation}
\Omega =\underset{n}{\oplus }e^{-\frac{1}{2}\int d\mu }\boldsymbol{1}_{n}
\label{2.10}
\end{equation}%
$\boldsymbol{1}_{n}$ denoting the identity function in the $n-$particle
subspace of a direct sum Hilbert space, the $\frac{1}{n!}$ factor in (\ref%
{2.8}) being recovered by the symmetrization operation.

However (\ref{2.7}) is not the most general consistent representation of the
nonrelativistic current algebra of a free boson gas, a more general one
being \cite{Goldin2}, the reducible functional%
\begin{equation}
L\left( f\right) =\int_{0}^{\infty }\exp \left( \rho \int \left( e^{if\left(
x\right) }-1\right) d^{3}x\right) d\xi \left( \rho \right)  \label{2.13}
\end{equation}%
with $\xi $ a positive measure on $\left[ 0,\infty \right) $ normalized so
that $\int_{0}^{\infty }d\xi \left( \rho \right) =1$. This
inifinite-dimensional compound Poisson measure may represent a boson gas
with density fluctuations. As pointed out in \cite{FracMinho}, among the
many possible reducible functionals consistent with (\ref{2.13}) there is a
fractional generalization of (\ref{2.8}), namely%
\begin{equation}
L_{\alpha }\left( f\right) =\sum_{n=0}^{\infty }\frac{E_{\alpha
}^{(n)}\left( -\int d\mu \right) }{n!}\left( \int e^{if\left( x\right) }d\mu
\right) ^{n}  \label{2.11}
\end{equation}%
($0<\alpha \leq 1$), which corresponds to a vacuum state%
\begin{equation}
\Omega _{\alpha }=\underset{n}{\oplus }\sqrt{E_{\alpha }^{(n)}\left( -\int
d\mu \right) }\boldsymbol{1}_{n}  \label{2.12}
\end{equation}%
$E_{\alpha }^{(n)}$ denoting the $n-$th derivative of the Mittag-Leffler
function \cite{Haubold}. $\Omega _{\alpha }$ differs from $\Omega $ in the
weight given to each one of the $n-$particle spaces. The measure associated
to the functional (\ref{2.11}) was called \textit{the infinite-dimensional
fractional Poisson measure} and the corresponding physical system \textit{%
the fractional boson gas}.

The reducible functional associated to the infinite-dimensional fractional
Poisson measure was introduced because the Mittag-Leffler is a very natural
analytic generalization of the exponential function. The main interest in
studying such an example is the possibility to analyse rigorously its
support properties as well as the Hilbert space structure, in particular the
nature of the $n$-particle subspaces. This is the mathematical motivation
for the study of the fractional boson gases. Of course it also suggests that
simlar support and Hilbert space modifications would occur for other
reducible functionals.

The study of the fractional boson gas has been carried out elsewhere \cite%
{FracMinho} and, for the convenience of the reader, the main results are
summarized in the Appendix. The meaning and relevance of the reducible
functionals of type (\ref{2.13}) becomes clear when finite temperature
functionals are computed. These were computed by Girard \cite{Girard} in the
current algebra framework.

\subsection{The zero temperature limit of finite-temperature functionals}

For $T\neq 0$, instead of the matrix element (\ref{2.6}), one computes%
\begin{equation}
L_{N,V}^{(T)}\left( f\right) =\frac{Tr\left( e^{-\beta H}e^{i\varrho \left(
f\right) }\right) }{Tr\left( e^{-\beta H}\right) }  \label{3.1}
\end{equation}%
for the canonical ensemble and%
\begin{equation}
L_{\mu ,V}^{(T)}\left( f\right) =\frac{Tr\left( e^{\beta \mu N}e^{-\beta
H}e^{i\varrho \left( f\right) }\right) }{Tr\left( e^{\beta \mu N}e^{-\beta
H}\right) }  \label{3.2}
\end{equation}%
for the grand canonical ensemble. $H$ is the free particle Hamiltonian, $%
\beta =\frac{1}{kT}$, $\mu $ is the chemical potential and $N$ the particle
number operator. Girard \cite{Girard} obtains for the grand canonical
functional%
\begin{equation}
L_{\mu ,V}^{(T)}\left( f\right) =\det [I-(e^{if(x)}-l)/(e^{\beta (H-\mu
)}-I)]^{-1}  \label{3.3}
\end{equation}%
However, taking the zero temperature limit of (\ref{3.3}) one does not
recover the infinite dimensional Poisson measure of (\ref{2.7}). Instead of (%
\ref{2.7}) the following functional is obtained \cite{Girard}%
\begin{equation}
L_{0}\left( \overline{\rho }\right) =\left( 1-\overline{\rho }\int \left(
e^{if\left( x\right) }-1\right) dx\right) ^{-1}  \label{3.4a}
\end{equation}%
which is seen to be a reducible functional, as in (\ref{2.13}), with density%
\begin{equation}
d\xi \left( \rho \right) =(\frac{1}{\overline{\rho }})e^{-\rho /\overline{%
\rho }}d\rho  \label{3.4b}
\end{equation}%
Physically this makes sense, because since the grand canonical ensemble only
fixes the particle number in average, it is reasonable that the
corresponding ground state be a state with density fluctuations. The zero
temperature limit of the grand canonical Boson gas is therefore a ground
state with density fluctuations defined by (\ref{3.4b}). A natural question
to ask is what is the physical meaning of all other reducible functionals.
One possible answer is the following result:

\textit{\# All reducible functionals of type (\ref{2.13})
(infinite-dimensional compound Poisson measures) may be obtained as
zero-temperature limits of superpositions of grand canonical free boson
gases with different chemical potentials}.

Consider a superposition of grand canonical free boson gases with different
chemical potentials, hence with different average densities $\overline{\rho }
$. Let the superposition be described by the measure $\nu \left( \overline{%
\rho }\right) $ with%
\begin{equation*}
\int d\overline{\rho }\nu \left( \overline{\rho }\right) =1
\end{equation*}%
Then the corresponding reducible functional would be%
\begin{equation*}
L_{\nu }\left( f\right) =\int_{0}^{\infty }\exp \left( \rho \int \left(
e^{if\left( x\right) }-1\right) d^{3}x\right) \Gamma \left( \rho \right)
d\rho
\end{equation*}%
with%
\begin{equation*}
\Gamma \left( \rho \right) =\int_{0}^{\infty }\nu \left( \overline{\rho }%
\right) (\frac{1}{\overline{\rho }})e^{-\rho /\overline{\rho }}d\overline{%
\rho }
\end{equation*}%
Changing variables $t=\frac{1}{\overline{\rho }}$%
\begin{equation*}
\Gamma \left( \rho \right) =\int_{0}^{\infty }\frac{\nu \left( \frac{1}{t}%
\right) }{t}e^{-\rho t}dt
\end{equation*}%
$\Gamma \left( \rho \right) $ is seen to be the Laplace transform of $\frac{%
\nu \left( \frac{1}{t}\right) }{t}$. Therefore, invertibility of the Laplace
transform implies that given a $\Gamma \left( \rho \right) $ one may find a $%
\nu \left( \overline{\rho }\right) -$superposition of grand canonical free
boson gases with that particular reducible functional.

This is one possible physical interpretation of the meaning of the reducible
functionals. Alternatively we might consider the reducible functionals in (%
\ref{2.13}) simply as zero-temperature limits of statistical ensembles with
density fluctuations. In favor of this alternative interpretation is the
fact that particles with different chemical potentials would be different
particles, but for example both the infinite dimensional Poisson measure and
the infinite dimensional fractional Poisson measure have the same support,
the configuration spaces of locally finite point measures without any
additional labelling (see the Appendix).

The study of the support of the measures associated to the irreducible and
the reducible cases gives some hints on their role as far as physical
modeling is concerned. For example, although the support for the infinite
dimensional Poisson measure and the fractional one (fractional boson gas)
are the same, the weights given to the $n-$particle states are different.
The grand canonical ensemble might not be the only useful particle number
fluctuation ensemble and different types of particle density fluctuations
might imply different low-temperature phase transition behaviors.

Here we explore this possibility by computing the modifications introduced
on the thermodynamic functions near the Bose-Einstein condensation
temperature when, instead of the usual grand canonical ensemble, we have
other types of particle number fluctuations, which would correspond in the
zero-temperature limit to general classes of reducible functionals. Based on
the equivalence result proved above this may be obtained by considering the
superposition of grand-canonical free boson gases with different chemical
potentials. For the grand-canonical free boson gas the number density $\frac{%
\left\langle N\right\rangle }{V}$ is%
\begin{equation*}
\frac{N}{V}=\frac{1}{V}\frac{z}{1-z}+\frac{1}{\lambda ^{3}}g_{3/2}\left(
z\right)
\end{equation*}%
the first term being the fraction of particles condensed in the ground
state, $z=e^{\beta \mu }$, $\lambda =\sqrt{\frac{2\pi \hslash ^{2}}{mkT}}$\
\ and $g_{3/2}\left( z\right) $ is the function%
\begin{equation*}
g_{3/2}\left( z\right) =\sum_{k=1}^{\infty }\frac{z^{k}}{k^{3/2}}
\end{equation*}%
For a superposition of grand-canonical free boson gases with different
chemical potentials we replace $e^{\beta \mu }$ by $e^{\beta \mu x}$ and
integrate over $x$ with a measure $\nu $ such that $\int dx\nu \left(
x\right) =1$. Then%
\begin{equation*}
\frac{N_{\nu }}{V}=\int_{0}^{\infty }dx\nu \left( x\right) \left\{ \frac{1}{V%
}\frac{z^{x}}{1-z^{x}}+\frac{1}{\lambda ^{3}}g_{3/2}\left( z^{x}\right)
\right\}
\end{equation*}%
Likewise the free energy becomes%
\begin{equation*}
\frac{U}{N}=\left\{ 
\begin{array}{llll}
\frac{3}{2}\frac{kTV}{N\lambda ^{3}}\int_{0}^{\infty }dx\nu \left( x\right)
g_{5/2}\left( z^{x}\right) &  &  & T>T_{c} \\ 
&  &  &  \\ 
\frac{3}{2}\frac{kTV}{N\lambda ^{3}}g_{5/2}\left( 1\right) &  &  & T<T_{c}%
\end{array}%
\right.
\end{equation*}%
with%
\begin{equation*}
g_{5/2}\left( z\right) =\sum_{k=1}^{\infty }\frac{z^{k}}{k^{5/2}}
\end{equation*}%
For $T>T_{c}$, $z\left( T\right) $ is obtained from%
\begin{equation*}
\int_{0}^{\infty }dx\nu \left( x\right) g_{3/2}\left( z^{x}\right) =\frac{%
N_{\nu }}{V}\left( \frac{2\pi \hslash ^{2}}{mk}\right) ^{3/2}T^{-3/2}
\end{equation*}%
and the specific heat $C_{V}=\frac{\partial U}{\partial T}$ for $T>T_{c}$%
\begin{eqnarray*}
C_{V} &=&\frac{15}{4}kVT^{3/2}\left( \frac{mk}{2\pi \hslash ^{2}}\right)
^{3/2}\int_{0}^{\infty }dx\nu \left( x\right) g_{5/2}\left( z^{x}\right) \\
&&+\frac{3}{2}\frac{kTV}{\lambda ^{3}}\int_{0}^{\infty }dx\nu \left(
x\right) g_{3/2}\left( z^{x}\right) xz^{-1}\frac{dz\left( T\right) }{dT}
\end{eqnarray*}

Let, as an example, $\nu \left( x\right) $ be a lognormal distribution
peaked at $x=1$%
\begin{equation*}
\nu \left( x\right) =\frac{1}{x\sigma \sqrt{2\pi }}e^{-\frac{\left( \ln
x-\sigma ^{2}\right) }{2\sigma ^{2}}}
\end{equation*}%
Computing $C_{V}$ from the equations above for several values of $\sigma $
one obtains the results plotted in the figure, where%
\begin{equation*}
T^{\ast }=\frac{mk}{2\pi \hslash ^{2}\rho ^{2/3}}T
\end{equation*}%
One sees that as $\sigma $ becomes larger the specific heat behavior, above
the condensation point, becomes sharper, more $\lambda $-like than the grand
canonical Bose condensation transition. Physically a larger $\sigma $ means
that the particle number fluctuations are larger than in the grand canonical
ensemble. Notice that this is a purely statistical effect associated to the
number fluctuations, no interaction being assumed in the Bose gas. 
\begin{figure}[htb]
\centering
\includegraphics[width=0.5\textwidth]{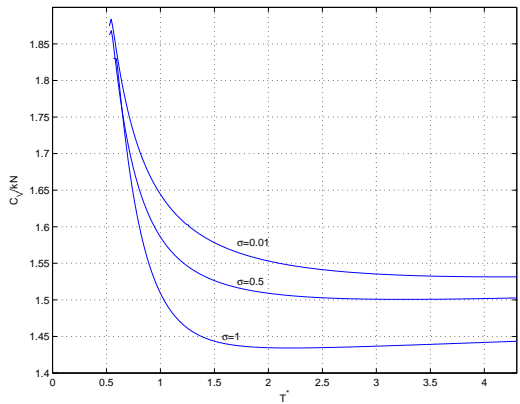}
\caption{Specific heat behavior, above the condensation point, for different
particle number fluctuations}
\label{CV}
\end{figure}

\section{Current algebra of many-body interacting systems}

\subsection{Representations of nonrelativistic currents, quasi-invariant
measures and the ground state formulation}

The search for representations of the current commutators (\ref{2.2}) or of
the semidirect product group generated by the exponentiated currents (\ref%
{2.3}) is a very general method to characterize many-body quantum systems.
In particular the current algebra structure is independent of whether one
deals with Boson or Fermi or even other exotic statistics.

If the relevant configuration space is $R^{d}$, a unitary representation of
the exponentiated currents in (\ref{2.3})%
\begin{eqnarray*}
U\left( f\right) &=&e^{i\varrho \left( f\right) } \\
V\left( \phi _{t}^{\mathbf{g}}\right) &=&e^{it\mathbf{J}\left( \mathbf{g}%
\right) }
\end{eqnarray*}%
is a unitary representation of a semidirect product of infinite dimensional
Lie groups%
\begin{equation}
G=\mathcal{D}\wedge Diff\left( \mathbb{R}^{d}\right)  \label{R2}
\end{equation}%
$\mathcal{D}$ being the commutative multiplicative group of Schwartz
functions $f\in \mathbb{C}_{0}^{\infty }\left( \mathbb{R}^{d}\right) $ and $%
Diff\left( \mathbb{R}^{d}\right) $ the group of smooth diffeomorphisms of $%
\mathbb{R}^{d}$. Of special concern here is the restriction to the connected
component of the identity $Diff_{0}\left( \mathbb{R}^{d}\right) $. The group
composition laws are%
\begin{eqnarray}
U\left( f_{1}\right) U\left( f_{2}\right) &=&U\left( f_{1}+f_{2}\right) 
\notag \\
V\left( \phi \right) U\left( f\right) &=&U\left( f\circ \phi \right) V\left(
\phi \right)  \notag \\
V\left( \phi _{1}\right) V\left( \phi _{2}\right) &=&V\left( \phi _{2}\circ
\phi _{1}\right)  \label{R3}
\end{eqnarray}

Taking the currents as the fundamental structures of quantum mechanics, all
physical models of (nonrelativistic) quantum mechanics should be obtained as
the unitary representations of the group $G$. A very general formulation for
the representations of this group starts from a space of square-integrable
functions $\mathcal{H}=\mathcal{L}_{\mu }^{2}\left( \Delta ,\mathcal{W}%
\right) $ where $\Delta $ is a configuration space, $\mathcal{W}$ an inner
product space and $\mu $ a measure on $\Delta $ quasi-invariant for the
diffeomorphisms action $V\left( \phi \right) $. Then%
\begin{equation}
\left( V\left( \phi \right) \Psi \right) \left( \gamma \right) =\chi _{\phi
}\left( \gamma \right) \Psi \left( \phi \gamma \right) \sqrt{\frac{d\mu
_{\phi }}{d\mu }\left( \gamma \right) }  \label{R30}
\end{equation}%
where $\gamma \in \Delta $, $\Psi \in \mathcal{H}$ and $\chi _{\phi }\left(
\gamma \right) :\mathcal{W\rightarrow W}$ is a family of unitary operators
in $\mathcal{W}$ satisfying the cocycle condition%
\begin{equation}
\chi _{\phi _{1}}\left( \gamma \right) \chi _{\phi _{2}}\left( \phi
_{1}\gamma \right) =\chi _{\phi _{1}\circ \phi _{2}}\left( \gamma \right)
\label{R3a}
\end{equation}%
Quasi-invariance of the measure $\mu $ is essential to insure the existence
of the Radon-Nikodym derivative in (\ref{R30}). On the other hand the
unitary operators $U\left( f\right) $ are assumed to act by multiplication%
\begin{equation}
\left( U\left( f\right) \Psi \right) \left( \gamma \right) =e^{i\left\langle
\gamma ,f\right\rangle }\Psi \left( \gamma \right)  \label{R3b}
\end{equation}%
the meaning of $\left\langle \gamma ,f\right\rangle $ depending on the
particular choice of configuration space.

A popular configuration space for statistical mechanics applications has
been the space of locally finite configurations, for which the Poisson
measure and some Gibbs measures have been extensively studied \cite%
{KondratRMP}. However other, more general, configuration spaces have been
proposed, the space of closed subsets of a manifold \cite{Ismagilov}, the
space of distributions $\mathcal{D}^{\prime }$ or $\mathcal{S}^{\prime }$,
the space of embeddings and immersions and the space of countable subsets of 
$\mathbb{R}^{d}$ (see for example \cite{Goldin3} \cite{Goldin-Moschella}).
Other configuration space worth to explore, when dealing with accumulation
points of infinite cardinality, is the space of ultradistributions or
ultradistributions of compact support, which have been found useful in
another context \cite{Vilela}.

The characterization of quantum systems through the construction of
quasi-invariant measures on configuration spaces is quite general. However,
a basic difficulty with this approach is that once a quasi-invariant measure
is obtained it might not be easy to figure out what is the physical
interaction (potential) that originates such measure. An alternative
constructive approach, already foreshadowed in \cite{Goldin2} \cite{Menik2},
is suggested by the following construction.

Let for definiteness the configuration space be $\mathcal{S}^{\prime }$ and
assume the representation to be cyclic, that is, there is a normalized
vector $\Omega \in \mathcal{H}$ such that the set $\left\{ U\left( f\right)
\Omega |f\in \mathcal{S}\right\} $ is dense in $\mathcal{H}$. Then the
functional%
\begin{equation}
L\left( f\right) =\left( \Omega ,U\left( f\right) \Omega \right)  \label{R3c}
\end{equation}%
with $L\left( 0\right) =1$ is positive definite and continuous. By the
Bochner-Minlos theorem it is the characteristic funtional of a measure on $%
\mathcal{H}$. The cyclic vector $\Omega $ becomes the central ingredient of
the construction and, as will be seen later on, it relates in an easy manner
to the interactions of the system.

In this spirit, Menikoff \cite{Menik2} proposed a set of axioms for the
construction of (nonrelativistic) quantum models: Let $\mathcal{H}$ be a
Hilbert space and $H$ a positive self-adjoint operator,

(i) There is a normalized state $\Omega $ of lowest energy. Then, by
eventually subtracting a constant from $H$%
\begin{equation}
H\Omega =0  \label{R4}
\end{equation}

(ii) $D=Span\left\{ U\left( f\right) \Omega ;f\in \mathcal{S}\right\} $ is
dense in $\mathcal{H}$ and $D$ is in the domain of $\mathcal{H}$.

(iii) Current conservation%
\begin{equation}
\left[ H,\rho \left( f\right) \right] =-i\mathbf{J}\left( \nabla f\right)
\label{R5}
\end{equation}

(iv) There is an antiunitary time reversal operator $\mathcal{T}$%
\begin{equation}
\mathcal{T}\rho \left( f\right) \mathcal{T}^{-1}=\rho \left( f\right) ;\;%
\mathcal{T}\text{\textbf{\ }}\mathbf{J}\left( \mathbf{g}\right) \mathcal{T}%
^{-1}=-\mathbf{J}\left( \mathbf{g}\right) ;\;\mathcal{T}\text{\textbf{\ }}%
\Omega =\Omega  \label{R6}
\end{equation}%
In this framework it is proved \cite{Menik2} that the matrix elements of $%
\mathbf{J}\left( \mathbf{g}\right) $ and $H$ are expressed in terms of those
of $\rho \left( f\right) $, namely%
\begin{eqnarray}
\left\langle e\left( f_{1}\right) \left\vert \mathbf{J}\left( \mathbf{g}%
\right) \right\vert e\left( f_{2}\right) \right\rangle &=&\frac{1}{2}%
\left\langle e\left( f_{1}\right) \left\vert \rho \left( \mathbf{g\cdot
\nabla }\left( f_{1}+f_{2}\right) \right) \right\vert e\left( f_{2}\right)
\right\rangle  \notag \\
\left\langle e\left( f_{1}\right) \left\vert H\right\vert e\left(
f_{2}\right) \right\rangle &=&\frac{1}{2}\left\langle e\left( f_{1}\right)
\left\vert \rho \left( \mathbf{\nabla }f_{1}\mathbf{\cdot \nabla }%
f_{2}\right) \right\vert e\left( f_{2}\right) \right\rangle  \label{R7}
\end{eqnarray}%
with $e\left( f\right) =\exp \left( i\rho \left( f\right) \Omega \right) $.
With time reversal invariance the Eqs. (\ref{R7}) follow from the
commutation relations%
\begin{eqnarray*}
\left[ \exp \left( i\rho \left( f\right) \right) ,\mathbf{J}\left( \mathbf{g}%
\right) \right] &=&-\rho \left( \mathbf{g\cdot \nabla f}\right) \exp \left(
i\rho \left( f\right) \right) \\
\left[ \exp \left( i\rho \left( f\right) \right) ,H\right] &=&\left( -%
\mathbf{J}\left( \nabla f\right) +\frac{1}{2}\rho \left( \nabla f\cdot
\nabla f\right) \right) \exp \left( i\rho \left( f\right) \right)
\end{eqnarray*}%
easily obtained from (\ref{2.2}). A Hermitian form on a dense set of states
does not uniquely determine an unbounded operator. Nevertheless, Eqs. (\ref%
{R7}) show the central role played by the density operator $\rho \left(
f\right) $ and the ground state $\Omega $ in the formulation of a quantum
theory. This information is summarized in the generating functional%
\begin{equation*}
L\left( f\right) =\left( \Omega ,U\left( f\right) \Omega \right)
\end{equation*}

Many-body quantum systems are usually explored by postulating a
interparticle potential and then obtaining the spectrum and eigenfunctions
of the corresponding Hamiltonian. What the above current algebra
characterization suggests is that a more natural (and complete)
specification of the system is through a guess to the ground state which may
be easier to infer from the macroscopic properties of the system than the
microscopic potential that leads to such behavior. The idea of \textit{%
"quantum mechanics from the ground state"} traces its origin to the papers
of Coester and Haag \cite{Coester} and Araki \cite{Araki}. It has been
further developed for single particle nonrelativistic quantum mechanics in
several papers \cite{Streit1} \cite{Streit2} \cite{Vilela-Vac}. In this
setting situations that would correspond to singular or nonlocal potentials
are easily handled. The current algebra formulation now suggests that such
an approach should also be carried out for many-body statistical mechanics.

Once a ground state function $\Omega =\exp \left( -W\right) $ without nodes
is defined, by adding a constant to the Hamiltonian%
\begin{equation*}
H=-\triangle +V
\end{equation*}%
such that%
\begin{equation*}
H\Omega =0
\end{equation*}%
the corresponding potential is%
\begin{equation*}
V=\frac{\triangle \Omega }{\Omega }=-\triangle W+\nabla W\bullet \nabla W
\end{equation*}%
Whereas in the approach through potentials, one usually restricts to a sum
of two body interactions, if an arbitrary ground state function is
postulated, it will in general correspond to potentials involving more than
two particles. Some exceptions are the harmonic interaction ground state in
arbitrary dimensions%
\begin{equation*}
W_{1}=\frac{\omega }{2}\sum_{i,j=1}^{N}\left( x_{i}-x_{j}\right) ^{2}
\end{equation*}%
and also%
\begin{equation*}
W_{2}=\frac{\omega }{2}\sum_{i,j=1}^{N}\left( x_{i}-x_{j}\right) ^{2}+\frac{%
\lambda }{2}\sum_{i\neq j}^{N}\log \left\vert x_{i}-x_{j}\right\vert
\end{equation*}%
in one dimension \cite{Calogero} \cite{Sutherland}.

In the following subsection one shows how the search for the ground
state,inspired by an algebraic structure of the currents may shed light on
the relevant\ physical problem of pairing in two-dimensional fermion systems.

\subsection{Hole pairing and current quivers}

Here, using currents, one attempts a formulation of a model for pairing as
is required in the high-temperature superconductor phenomenon. First a short
outline of the most relevant phenomenological facts which inspire the search
for the elements of the model.

\textit{First: The role of hole carriers and antiferromagnetic interactions}

Experiments have shown that the charge carriers have hole character for all
classes of high-temperature superconductors and the superconducting state
arises near the antiferromagnetic phase, experiments on the inelastic
magnetic scattering of neutrons indicating the existence of strong magnetic
fluctuations in the doped region, even beyond the limits of the
antiferromagnetic phase. Though the long-range order disappears in the
metallic and the superconducting phases, strong fluctuations with a wide
spectrum of excitations are conserved, suggesting at least some local
antiferromagnetic order. The closeness of the superconducting to the
antiferromagnetic transition emphasizes the important role of spin
fluctuations.

\textit{Second: The dual role of a gap and the phase coherence}

In high-temperature superconductors, a gap is present even in the absence of
phase coherence, i.e. in nonsuperconducting specimens. It appears at
temperatures less than some characteristic temperature which depends on the
doping. The (pseudo)gap is related to the appearance of coupled pairs, even
before the onset of\ the phase coherence responsible for the change of the
resistance.

Therefore a key question is the nature of the mechanism of pairing of the
carriers. Many different models were proposed, among which the following
ones: the magnon model, the exciton model, the resonant valence bond,
bipolaronic model, bisoliton model, anharmonic model, local pairs model,
plasmon model, etc. All these models use the concept of pairing with the
subsequent formation of a Bose-condensation at the superconducting
transition. Pairing is therefore the central physical mechanism to be
explained.

All this experimental information led to relate high-temperature
superconductivity to the class of strongly correlated systems, the Hubbard
model, the $t-J$ model, the antiferromagnetic Heisenberg model, etc. At the
basis of these models are two simple ideas: first that in a regular array of
lattice positions, the dominant positive contribution to the energy is the
Coulomb repulsion when two (opposite spin) electrons occupy the same site,
modelled by a term%
\begin{equation}
\sum_{a}c_{a\uparrow }^{\dag }c_{a\uparrow }c_{a\downarrow }^{\dag
}c_{a\downarrow }  \label{HP1a}
\end{equation}%
and second that the energy decreases when the electrons are allowed to hop
between closeby sites, modelled by%
\begin{equation}
-\sum_{\left( a,b\right) ,\sigma }c_{a\sigma }^{\dag }c_{b\sigma }
\label{HP1b}
\end{equation}%
$\left( a,b\right) $ meaning closeby sites, nearest-neighbors or
next-to-nearest neighbors. $c_{a\sigma }$ and $c_{a\sigma }^{\dag }$
destruction and creation electron operators at the site $a$ and $\sigma $ is
the spin orientation $\left( \uparrow ,\downarrow \right) $. The interacting
terms (\ref{HP1a}) and (\ref{HP1b}) form the basis of the Hubbard model.
However, there is some evidence (see for example \cite{Vilela1}) that by
itself the Hubbard model is not sufficient to provide an hole pairing
mechanism and that extra interactions must be called into play. We discuss
this matter in terms of currents.

The interaction terms may be expressed in physical variables, that is,
currents and densities. Notice however that the most appropriate algebraic
structure for these physical variables might not be a Lie algebra. Consider
a 2D square lattice with the atoms at the lattice vertices. The physical
variables are the densities at each site $a$ 
\begin{equation}
\rho _{\sigma }\left( a\right) =c_{a\sigma }^{\dag }c_{a\sigma }  \label{HP2}
\end{equation}%
and the currents%
\begin{equation}
J_{\sigma }\left( a,b\right) =-i\left( c_{b\sigma }^{\dag }c_{a\sigma
}-c_{a\sigma }^{\dag }c_{b\sigma }\right)  \label{HP3}
\end{equation}%
corresponding to electron fluxes between the sites $a$ and $b$. The
commutation relations are%
\begin{eqnarray}
\left[ \rho _{\sigma }\left( a\right) ,J_{\sigma ^{\prime }}\left(
m,n\right) \right] &=&-i\left( \delta _{a,n}-\delta _{a,m}\right) K\left(
m,n\right) \delta _{\sigma \sigma ^{\prime }}  \notag \\
\left[ \rho _{\sigma }\left( a\right) ,K_{\sigma ^{\prime }}\left(
m,n\right) \right] &=&i\left( \delta _{a,n}-\delta _{a,m}\right) J\left(
m,n\right) \delta _{\sigma \sigma ^{\prime }}  \notag \\
\left[ J_{\sigma }\left( a,b\right) ,J_{\sigma ^{\prime }}\left( m,n\right) %
\right] &=&i\left( -\delta _{a,m}J\left( b,n\right) +\delta _{a,n}J\left(
b,m\right) -\delta _{b,n}J\left( a,m\right) +\delta _{b,m}J\left( a,n\right)
\right) \delta _{\sigma \sigma ^{\prime }}  \notag \\
\left[ J_{\sigma }\left( a,b\right) ,K_{\sigma ^{\prime }}\left( m,n\right) %
\right] &=&i\left( -\delta _{a,m}K\left( n,b\right) -\delta _{a,n}K\left(
m,b\right) +\delta _{b,n}K\left( m,a\right) +\delta _{b,m}K\left( n,a\right)
\right) \delta _{\sigma \sigma ^{\prime }}  \notag \\
\left[ K_{\sigma }\left( a,b\right) ,K_{\sigma ^{\prime }}\left( m,n\right) %
\right] &=&i\left( \delta _{a,m}J\left( n,b\right) +\delta _{a,n}J\left(
m,b\right) +\delta _{b,n}J\left( m,a\right) +\delta _{b,m}J\left( n,a\right)
\right) \delta _{\sigma \sigma ^{\prime }}  \label{HP4}
\end{eqnarray}%
$K_{\sigma }\left( m,n\right) $ being the operator%
\begin{equation}
K_{\sigma }\left( m,n\right) =c_{n\sigma }^{\dag }c_{m\sigma }+c_{m\sigma
}^{\dag }c_{n\sigma }  \label{HP5}
\end{equation}%
This is the operator that in the continuum case leads to the term $\varrho
\left( \mathbf{g\bullet \nabla }f\right) $ in right hand side of (\ref{2.2}%
). Notice that $K_{\sigma }\left( m,m\right) \equiv 2\rho _{\sigma }\left(
m\right) $.

\begin{figure}[htb]
\centering
\includegraphics[width=0.5\textwidth]{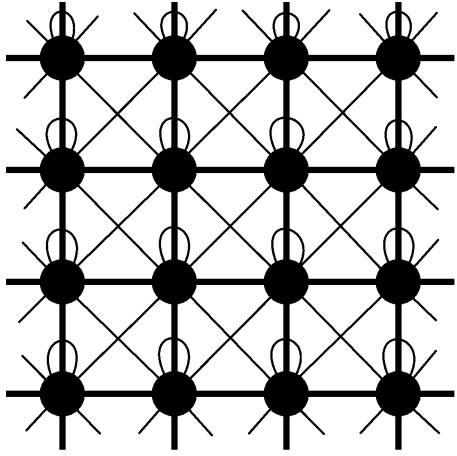}
\caption{A current quiver}
\label{quiver}
\end{figure}

One sees from the commutation relation of the currents that starting from
currents connecting close neighbors one obtains, by successive commutators,
currents involving direct hoppings between all sites in the lattice which,
for the strongly correlated systems, are of no immediate physical interest.
Therefore a Lie algebra is not an useful algebraic structure for these
currents. Instead, restricting to nearest neighbor and next-to-nearest
neighbor hoppings one obtains the following quiver (Fig.\ref{quiver}). 
A quiver is a directed graph. A representation of a quiver assigns a vector
space $\mathcal{N}$ to each vertex, and a linear map to each edge (arrow).
In the \textit{current quiver} of Fig.\ref{quiver} the arrows connecting the
vertices to themselves are charge density contributions $\rho _{\sigma
}\left( a\right) $ and those connecting different vertices correspond to the
operators%
\begin{equation}
V_{\sigma }\left( a,b\right) =\frac{1}{2}\left( K_{\sigma }\left( a,b\right)
+iJ_{\sigma }\left( a,b\right) \right)  \label{HP5a}
\end{equation}%
$V_{\sigma }\left( a,b\right) $ being a directed map corresponding to an hop
from site $a$ to site $b$. Notice that $V_{\sigma }\left( a,a\right) \equiv
\rho _{\sigma }\left( a\right) $.

To each vertex one assigns a four-dimensional space corresponding to the
electron configurations ($\uparrow \downarrow $,$\uparrow $,$\downarrow $,$%
\bigcirc $), respectively double occupancy, spin up, spin down and a hole.
The directed hop maps $V_{\sigma }\left( a,b\right) $ are represented by $%
4\times 4$ matrices with elements%
\begin{equation}
V_{\uparrow }\left( a,b\right) =\left( 
\begin{array}{cccc}
0 & 0 & 1 & 1 \\ 
0 & 0 & 1 & 1 \\ 
0 & 0 & 0 & 0 \\ 
0 & 0 & 0 & 0%
\end{array}%
\right) ;V_{\downarrow }\left( a,b\right) =\left( 
\begin{array}{cccc}
0 & 1 & 0 & 1 \\ 
0 & 0 & 0 & 0 \\ 
0 & 1 & 0 & 1 \\ 
0 & 0 & 0 & 0%
\end{array}%
\right)  \label{HP6}
\end{equation}%
each element of the matrices accounting for the possible hopping
contributions from vertex $a$ to $b$. For the arrows connecting one vertex
to itself, the representation maps are%
\begin{equation}
\rho _{\uparrow }\left( a\right) =\left( 
\begin{array}{cccc}
1 & 0 & 0 & 0 \\ 
0 & 1 & 0 & 0 \\ 
0 & 0 & 0 & 0 \\ 
0 & 0 & 0 & 0%
\end{array}%
\right) ;\rho _{\downarrow }\left( a\right) =\left( 
\begin{array}{cccc}
1 & 0 & 0 & 0 \\ 
0 & 0 & 0 & 0 \\ 
0 & 0 & 1 & 0 \\ 
0 & 0 & 0 & 0%
\end{array}%
\right)  \label{HP7}
\end{equation}%
The relevant algebraic framework is therefore the quiver with maps $\rho
_{\sigma }\left( a\right) $ and $V_{\sigma }\left( a,b\right) $ and
composition laws%
\begin{equation}
V_{\sigma }\left( a,b\right) V_{\sigma }\left( m,n\right) =\delta
_{an}V_{\sigma }\left( m,b\right) +\delta _{mn}V\left( a,b\right) -V_{\sigma
}\left( m,b\right) V_{\sigma }\left( a,n\right)  \label{HP8}
\end{equation}%
In particular from $V_{\sigma }\left( a,a\right) \equiv \rho _{\sigma
}\left( a\right) $ it follows that%
\begin{equation}
V_{\sigma }\left( a,b\right) V_{\sigma }\left( b,a\right) =\rho _{\sigma
}\left( b\right) \left( 1-\rho _{\sigma }\left( a\right) \right)  \label{HP9}
\end{equation}

The state $\Psi $ of the system is the tensor product of the states $\psi
_{i}\in \mathcal{N}$ for each vertex. Stationary states of the quiver are
states that are invariant for some iteration of the quiver. Collecting the
simplest quiver operations that leave a state $\Psi $ invariant, a general
form for the stationary energy associated to the quiver is%
\begin{eqnarray}
E &=&U\sum_{a}\rho _{\uparrow }\left( a\right) \rho _{\downarrow }\left(
a\right) -t\sum_{\left\langle a,b\right\rangle ,\sigma }V_{\sigma }\left(
a,b\right) V_{\sigma }\left( b,a\right)  \notag \\
&&+k\sum_{\left\langle a,b\right\rangle ,\sigma }\left( 1-\rho _{\sigma
}\left( a\right) \right) \left( 1-\rho _{-\sigma }\left( a\right) \right)
\left( 1-\rho _{\sigma }\left( b\right) \right) \left( 1-\rho _{-\sigma
}\left( b\right) \right)  \notag \\
&&-J\sum_{a,\sigma }\left\{ \alpha +\beta \left( 1-\rho _{\sigma }\left(
a\right) \right) \left( 1-\rho _{-\sigma }\left( a\right) \right) \right\}
\sum_{\left[ n_{a}n_{\sigma }^{\prime }\right] ,\sigma ^{\prime }}V_{\sigma
^{\prime }}\left( n_{a},n_{a}^{\prime }\right) V_{\sigma ^{\prime }}\left(
n_{a}^{\prime },n_{a}\right)  \notag \\
&&  \label{HP10}
\end{eqnarray}%
where $\left\langle a,b\right\rangle $ denotes nearest-neighbors and $\left[
a,b\right] $ next-to-nearest-neighbors.

The first term is a positive contribution from Coulomb repulsion of two
electrons in the same lattice site. The second is a symmetric hopping term
between nearest-neighbor sites. The third is a hole repulsion term for holes
at nearest-neighbor sites and finally the last term accounts for the hopping
contributions between the neighbors of site $a$, which in the square lattice
are next-to-nearest neighbors. In this last term two main possibilities are
considered. If $\alpha =1$ and $\beta =0$ one has an unconditional
next-to-nearest hopping contribution of intensity $-J$. However, if $\alpha
=0$ and $\beta =1$, hopping between the neighbors of site $a$ only takes
place if there is an hole in this site. The physical idea behind this
possibility is that the hole distorts the orbitals in its neighborhood
increasing the overlap of the wave functions of its neighbors.

Notice that our definition of the quiver energy does differ from similar
operators derived from the Hubbard model by canonical transformations and
leading order truncations. Using (\ref{HP9}) the quiver energy is rewritten 
\begin{eqnarray*}
E &=&U\sum_{a}\rho _{\uparrow }\left( a\right) \rho _{\downarrow }\left(
a\right) -t\sum_{\left\langle a,b\right\rangle ,\sigma }\rho _{\sigma
}\left( b\right) \left( 1-\rho _{\sigma }\left( a\right) \right) \\
&&+k\sum_{\left\langle a,b\right\rangle ,\sigma }\left( 1-\rho _{\sigma
}\left( a\right) \right) \left( 1-\rho _{-\sigma }\left( a\right) \right)
\left( 1-\rho _{\sigma }\left( b\right) \right) \left( 1-\rho _{-\sigma
}\left( b\right) \right) \\
&&-J\sum_{a,\sigma }\left\{ \alpha +\beta \left( 1-\rho _{\sigma }\left(
a\right) \right) \left( 1-\rho _{-\sigma }\left( a\right) \right) \right\}
\sum_{\left[ n_{a}n_{\sigma }^{\prime }\right] ,\sigma ^{\prime }}\rho
_{\sigma ^{\prime }}\left( n_{a}^{\prime }\right) \left( 1-\rho _{\sigma
^{\prime }}\left( n_{a}\right) \right)
\end{eqnarray*}%
Having the stationary energy fully expressed in number operators, the search
for minimum energy states becomes a simple counting matter. Consider a 2D
square lattice with N sites, $N-H$ electrons and $H$ holes ($H\ll N$) and%
\begin{equation*}
U\gg t,\;t>J,\;4J>k
\end{equation*}%
To lower the energy, the large $U$ value implies single occupancy of the
lattice sites and $t>J$ (local) antiferromagnetic order. In the case $\alpha
=1$, $\beta =0$ a lowest energy estimate yields%
\begin{equation*}
E_{1,0}\simeq -t\frac{\left( N-H\right) \left( N-H-1\right) }{2}-4JH
\end{equation*}%
the holes being spread over the lattice without any special correlation
among them. Any hole pairing would imply a $k$ positive contribution. In
contrast for the case $\alpha =0$, $\beta =1$ the minimum energy estimate is%
\begin{equation*}
E_{0,1}\simeq -t\frac{\left( N-H\right) \left( N-H-1\right) }{2}-2JH+k\frac{H%
}{2}
\end{equation*}%
The physical mechanism is clear. Although the wave function overlap in the
neighborhood of an hole facilitates hopping between the neighbors of the
hole, the local antiferromagnetic order frustrates this hopping. Hence, to
lower the energy, another hole must be attracted to the neighborhood of the
first hole and all holes are paired. Larger hole clusters will be avoided if 
$2k>5J$.

Hole pairing is a precondition to the latter formation of the coherent state
leading to superconductivity. The hole-induced hopping described here is a
plausible mechanism for a possible hole pairing mechanism.

\section{Conclusions}

1 - In contrast with the quantum fields of canonical quantization, local
currents are directly related to physical observables. In addition, whereas
there are strong uniqueness results for the representation of
finite-dimensional canonical commutation relations, the algebra of
nonrelativistic currents has many non-equivalent represenations, each
particular physical system corresponding to a different one. These two facts
make (non-relativistic) current algebra a candidate of choice for the
formulation of the statistical mechanics of many-body systems.

2 - The construction of representations of the current algebra may be
carried out either by defining quasi-invariant measures on configuration
spaces or by a generating functional obtained from a (ground state) cyclic
vector. It is argued in this paper that the second approach is more
appropriate as a modeling tool for physical systems. An extensive
application of this approach was done in the construction of a hole pairing
model. It has also been found that for some models, instead of the full
current algebra, a subset of operators is sufficient. A current quiver is
used in the hole pairing model.

3 - For boson systems, in addition to the ground state of the fixed density
N/V limit, other reducible functionals might be useful to describe systems
with number density fluctuations. A reducible functional is already implicit
in the use of the grand canonical ensemble, but other functionals provide
alternative phase transition behaviors.

\section*{Appendix: The support of the infinite-dimensional Poisson and
fractional Poisson measures}

Here, for the reader's convenience and in particular as a background to
Section 2, a short summary is given of the properties of the
infinite-dimensional Poisson measure, its support on configuration spaces 
\cite{Albeverio1} \cite{Albeverio2} \cite{Kondratiev1} \cite{Kondratiev2} 
\cite{Kondratiev3} \cite{Oliveira1} \cite{Kuna} as well as of a fractional
generalization \cite{Oliveira2} \cite{FracMinho}.

\subsection*{The infinite-dimensional Poisson measure}

The Poisson measure $\pi $ in $\mathbb{R}$\ (or $\mathbb{N}$) is%
\begin{equation}
\pi \left( A\right) =e^{-s}\sum_{n\in A}\frac{s^{n}}{n!}  \label{A.1}
\end{equation}%
the parameter $s$ being called the \textit{intensity}. The Laplace transform
of $\pi $ is%
\begin{equation*}
l_{\pi }\left( \lambda \right) =\mathbb{E}\left( e^{\lambda \cdot }\right)
=e^{-s}\sum_{n=0}^{\infty }\frac{s^{n}}{n!}e^{\lambda n}=e^{s\left(
e^{\lambda }-1\right) }
\end{equation*}%
and for $n$-tuples of independent Poisson variables one would have the
Laplace transform%
\begin{equation*}
l_{\pi }\left( \boldsymbol{\lambda }\right) =e^{\sum s_{k}\left( e^{\lambda
_{k}}-1\right) },\quad \boldsymbol{\lambda }=(\lambda _{1},\ldots ,\lambda
_{n})
\end{equation*}%
Continuing $\lambda _{k}$ to imaginary arguments $\lambda _{k}=if_{k}$,
yields the characteristic function,%
\begin{equation}
C_{\pi }\left( \lambda \right) =e^{\sum s_{k}\left( e^{if_{k}}-1\right) }
\label{A.2}
\end{equation}%
An infinite-dimensional generalization is obtained by generalizing (\ref{A.2}%
) to%
\begin{equation}
C\left( \varphi \right) =e^{\int \left( e^{i\varphi \left( x\right)
}-1\right) \,d\mu \left( x\right) }  \label{A.3}
\end{equation}%
for test functions $\varphi \in \mathcal{D}\left( M\right) $, $M$ being the
space of $C^{\infty }$-functions of compact support in a manifold $M$. It is
easy to prove, using the Bochner-Minlos theorem, that $C\left( \varphi
\right) $ is indeed the Fourier transform of a measure on the distribution
space $\mathcal{D}^{\prime }\left( M\right) $.

A support for this measure is obtained in the space of locally finite
subsets. The \textit{configuration space} $\Gamma :=\Gamma _{M}$ over the
manifold $M$ is defined as the set of all locally finite subsets of $M$
(simple configurations)%
\begin{equation}
\Gamma :=\{\gamma \subset M\,:\,|\gamma \cap K|<\infty \;\mathrm{\
for\;any\;compact\;}K\subset M\}  \label{A.4}
\end{equation}%
Here $|A|$ denotes the cardinality of the set $A$. As usual one identifies
each $\gamma \in \Gamma $ with a non-negative integer-valued Radon measure,%
\begin{equation*}
\Gamma \ni \gamma \mapsto \sum_{x\in \gamma }\delta _{x}\in \mathcal{M}(M)
\end{equation*}%
where $\delta _{x}$ is the Dirac measure with unit mass at $x$ and $\mathcal{%
M}(M)$ denotes the set of all non-negative Radon measures on $M$. In this
way the space $\Gamma $ can be endowed with the relative topology as a
subset of the space of measures $\mathcal{M}(M)$ with the vague topology,
i.e., the weakest topology on $\Gamma $ for which the mappings%
\begin{equation*}
\Gamma \ni \gamma \mapsto \left\langle \gamma ,f\right\rangle
:=\int_{M}f(x)d\gamma (x)=\sum_{x\in \gamma }f(x)
\end{equation*}%
are continuous for all real-valued continuous functions $f$ on $M$ with
compact support. Denote the corresponding Borel $\sigma -$algebra on $\Gamma 
$ by $\mathcal{B}\left( \Gamma \right) $.

For each $Y\in \mathcal{B}(M)$ let us consider the space $\Gamma _{Y}$ of
all configurations contained in $Y$, $\Gamma _{Y}:=\left\{ \gamma \in \Gamma
:\left\vert \gamma \cap (X\backslash Y)\right\vert =0\right\} $, and the
space $\Gamma _{Y}^{(n)}$ of $n$-point configurations,%
\begin{equation*}
\Gamma _{Y}^{(n)}:=\left\{ \gamma \in \Gamma _{Y}:\left\vert \gamma
\right\vert =n\right\} ,n\in \mathbb{N},\quad \Gamma _{Y}^{(0)}:=\left\{
\emptyset \right\}
\end{equation*}

A topological structure may be introduced on $\Gamma _{Y}^{(n)}$ through the
natural surjective mapping of $\widetilde{Y^{n}}:=\left\{
(x_{1},...,x_{n}):x_{i}\in Y,x_{i}\neq x_{j}\text{ if }i\neq j\right\} $
onto $\Gamma _{Y}^{(n)}$,%
\begin{equation*}
\begin{array}{ll}
\mathrm{sym}_{Y}^{n}:\widetilde{Y^{n}} & \longrightarrow \Gamma _{Y}^{(n)}
\\ 
(x_{1},...,x_{n}) & \longmapsto \left\{ x_{1},...,x_{n}\right\}%
\end{array}%
\end{equation*}%
which is at the origin of a bijection between $\Gamma _{Y}^{(n)}$ and the
symmetrization $\widetilde{Y^{n}}/S_{n}$ of $\widetilde{Y^{n}}$, $S_{n}$
being the permutation group over $\left\{ 1,...,n\right\} $. Thus, $\mathrm{%
sym}_{Y}^{n}$ induces a metric on $\Gamma _{Y}^{(n)}$ and the corresponding
Borel $\sigma -$algebra $\mathcal{B}\left( \Gamma _{Y}^{(n)}\right) $ on $%
\Gamma _{Y}^{(n)}$.

For $\Lambda \in \mathcal{B}(M)$ with compact closure ($\Lambda \in \mathcal{%
B}_{c}(M)$), it clearly follows from (\ref{A.4}) that%
\begin{equation*}
\Gamma _{\Lambda }=\bigsqcup_{n=0}^{\infty }\Gamma _{\Lambda }^{(n)}
\end{equation*}%
the $\sigma $-algebra $\mathcal{B}(\Gamma _{\Lambda })$ being defined by the
disjoint union of the $\sigma -$algebras $\mathcal{B}\left( \Gamma
_{Y}^{(n)}\right) $, $n\in \mathbb{N}_{0}$.

For each $\Lambda \in \mathcal{B}_{c}(M)$ there is a natural measurable
mapping $p_{\Lambda }:\Gamma \rightarrow \Gamma _{\Lambda }$. Similarly,
given any pair $\Lambda _{1},\Lambda _{2}\in \mathcal{B}_{c}(M)$ with $%
\Lambda _{1}\subset \Lambda _{2}$ there is a natural mapping $p_{\Lambda
_{2},\Lambda _{1}}:\Gamma _{\Lambda _{2}}\rightarrow \Gamma _{\Lambda _{1}}$%
. They are defined, respectively, by

\begin{equation*}
\begin{array}{ll}
p_{\Lambda }: & \Gamma \longrightarrow \Gamma _{\Lambda } \\ 
& \gamma \longmapsto \gamma _{\Lambda }:=\gamma \cap \Lambda%
\end{array}%
\quad 
\begin{array}{ll}
p_{\Lambda _{2},\Lambda _{1}}: & \Gamma _{\Lambda _{2}}\longrightarrow
\Gamma _{\Lambda _{1}} \\ 
& \gamma \longmapsto \gamma _{\Lambda _{1}}%
\end{array}%
\end{equation*}%
It can be shown that $(\Gamma ,\mathcal{B}(\Gamma ))$ coincides (up to an
isomorphism) with the projective limit of the measurable spaces $(\Gamma
_{\Lambda },\mathcal{B}(\Gamma _{\Lambda }))$, $\Lambda \in \mathcal{B}%
_{c}\left( M\right) $, with respect to the projection $p_{\Lambda }$, i.e., $%
\mathcal{B}(\Gamma )$ is the smallest $\sigma -$algebra on $\Gamma $ with
respect to which all projections $p_{\Lambda }$, $\Lambda \in \mathcal{B}%
_{c}\left( M\right) $, are measurable.

Let now $\mu $ be a measure on the underlying measurable space $(M,\mathcal{B%
}(M))$ and consider for each $n\in \mathbb{N}$ the product measure $\mu
^{\otimes n}$ on $(M^{n},\mathcal{B}(M^{n}))$. Since $\mu ^{\otimes
n}(M^{n}\backslash \widetilde{M^{n}})=0$, one may consider for each $\Lambda
\in \mathcal{B}_{c}(M)$ the restriction of $\mu ^{\otimes }$ to $(\widetilde{%
\Lambda ^{n}},\mathcal{B}(\widetilde{\Lambda ^{n}}))$, which is a finite
measure, and then the image measure $\mu _{\Lambda }^{(n)}$ on $(\Gamma
_{\Lambda }^{(n)},\mathcal{B}(\Gamma _{\Lambda }^{(n)}))$ under the mapping $%
\mathrm{sym}_{\Lambda }^{n}$,%
\begin{equation*}
\mu _{\Lambda }^{(n)}:=\mu ^{\otimes n}\circ (\mathrm{sym}_{\Lambda
}^{n})^{-1}
\end{equation*}%
For $n=0$ we set $\mu _{\Lambda }^{(0)}:=1$. Now, one may define a
probability measure $\pi _{\mu ,\Lambda }$ on $(\Gamma _{\Lambda },\mathcal{B%
}(\Gamma _{\Lambda }))$ by%
\begin{equation}
\pi _{\mu ,\Lambda }:=\sum_{n=0}^{\infty }\frac{\exp (-\mu (\Lambda ))}{n!}%
\mu _{\Lambda }^{(n)}  \label{A.5}
\end{equation}%
The family $\{\pi _{\mu ,\Lambda }:\Lambda \in \mathcal{B}_{c}(M)\}$ of
probability measures yields a probability measure on $(\Gamma ,\mathcal{B}%
(\Gamma ))$ with the $\pi _{\mu ,\Lambda }$ as projections. This family is
consistent, that is,%
\begin{equation*}
\pi _{\mu ,\Lambda _{1}}=\pi _{\mu ,\Lambda _{2}}\circ p_{\Lambda
_{2},\Lambda _{1}}^{-1},\quad \forall \,\Lambda _{1},\Lambda _{2}\in 
\mathcal{B}_{c}(M),\Lambda _{1}\subset \Lambda _{2}
\end{equation*}%
and thus, by the version of Kolmogorov's theorem for the projective limit
space $(\Gamma ,\mathcal{B}(\Gamma ))$, the family $\{\pi _{\mu ,\Lambda
}:\Lambda \in \mathcal{B}_{c}(M)\}$ determines uniquely a measure $\pi _{\mu
}$ on $(\Gamma ,\mathcal{B}(\Gamma ))$ such that%
\begin{equation*}
\pi _{\mu ,\Lambda }=\pi _{\mu }\circ p_{\Lambda }^{-1},\quad \forall
\,\Lambda \in \mathcal{B}_{c}(M)
\end{equation*}

The next step is to compute the characteristic functional of the measure $%
\pi _{\mu }$. Given a $\varphi \in \mathcal{D}(M)$ we have supp$\,\varphi
\subset \Lambda $ for some $\Lambda \in \mathcal{B}_{c}(M)$, meaning that%
\begin{equation*}
\langle \gamma ,\varphi \rangle =\langle p_{\Lambda }(\gamma ),\varphi
\rangle ,\quad \forall \,\gamma \in \Gamma
\end{equation*}%
Thus%
\begin{equation*}
\int_{\Gamma }e^{i\langle \gamma ,\varphi \rangle }d\pi _{\mu }(\gamma
)=\int_{\Gamma _{\Lambda }}e^{i\langle \gamma ,\varphi \rangle }d\pi _{\mu
,\Lambda }(\gamma )
\end{equation*}%
and the definition (\ref{A.5}) of the measure $\pi _{\mu ,\Lambda }$ yields
for the right-hand side of the equality%
\begin{equation*}
\sum_{n=0}^{\infty }\frac{\exp (-\mu (\Lambda ))}{n!}\int_{\Lambda
^{n}}e^{i(\varphi (x_{1})+\ldots +\varphi (x_{n}))}d\mu ^{\otimes
n}(x)=\sum_{n=0}^{\infty }\frac{\exp (-\mu (\Lambda ))}{n!}\left(
\int_{\Lambda }e^{i\varphi (x)}d\mu (x)\right) ^{n}
\end{equation*}%
which corresponds to the Taylor expansion of the characteristic function (%
\ref{A.3}) of the infinite-dimensional Poisson measure%
\begin{equation*}
\exp \left( \int_{\Lambda }(e^{i\varphi (x)}-1)\,d\mu (x)\right)
\end{equation*}%
This shows that the probability measure on $(\mathcal{D}^{^{\prime }}(M),%
\mathcal{C}_{\sigma }(\mathcal{D}^{^{\prime }}(M)))$ given by (\ref{A.3}) is
actually supported on generalized functions of the form $\sum_{x\in \gamma
}\delta _{x}$, $\gamma \in \Gamma $. Thus, the inifinite-dimensional Poisson
measure $\pi _{\mu }$ can either be considered as a measure on $(\Gamma ,%
\mathcal{B}(\Gamma ))$ or on $(\mathcal{D}^{\prime },\mathcal{C}_{\sigma }(%
\mathcal{D}^{\prime }(M)))$. Notice that, in contrast to $\Gamma $, $%
\mathcal{D}^{\prime }(M)\supset \Gamma $ is a linear space. Since $\pi _{\mu
}(\Gamma )=1$, the measure space $(\mathcal{D}^{\prime }(M),\mathcal{C}%
_{\sigma }(\mathcal{D}^{\prime }(M)),\pi _{\mu })$ can, in this way, be
regarded as a linear extension of the Poisson space $(\Gamma ,\mathcal{B}%
(\Gamma ),\pi _{\mu })$.

\subsection*{The infinite-dimensional fractional Poisson measure}

The Poisson process has a fractional generalization \cite{Mainardi} \cite%
{Vilela2}, the probability of $n$ events being

\begin{equation}
P\left( X=n\right) =\frac{s^{\alpha n}}{n!}E_{\alpha }^{(n)}\left(
-s^{\alpha }\right)  \label{B.1}
\end{equation}%
$E_{\alpha }^{(n)}$ denoting the $n$-th derivative of the Mittag-Leffler
function.%
\begin{equation}
E_{\alpha }\left( z\right) =\sum_{n=0}^{\infty }\frac{z^{n}}{\Gamma \left(
\alpha n+1\right) },\quad z\in \mathbb{C}  \label{B.2}
\end{equation}%
$\left( \alpha >0\right) $. In contrast with the Poisson case ($\alpha =1$),
this process has power law asymptotics rather than exponential, which
implies that it is not longer Markovian. The characteristic function of this
process is given by%
\begin{equation}
C_{\alpha }\left( \lambda \right) =E_{\alpha }\left( s^{\alpha }\left(
e^{i\lambda }-1\right) \right)  \label{B.3}
\end{equation}%
By analogy with (\ref{A.3}) an infinite-dimensional generalization is
obtained by generalizing (\ref{B.3}) to%
\begin{equation}
C_{\alpha }\left( \varphi \right) :=E_{\alpha }\left( \int (e^{i\varphi
\left( x\right) }-1)\,d\mu \left( x\right) \right) ,\quad \varphi \in 
\mathcal{D}\left( M\right)  \label{B.4}
\end{equation}%
with $\mu $ a positive intensity measure fixed on the underlying manifold $M$%
. Using the Bochner-Minlos theorem and the complete monotonicity of the
Mittag-Leffler function $C_{\alpha }$ is shown \cite{Oliveira2} to be the
characteristic functional of a probability measure $\pi _{\mu }^{\alpha }$
on the distribution space $\mathcal{D}^{\prime }\left( M\right) $

It turns out that this measure is also supported in configuration spaces and
the formulation in configuration spaces provides, through the Kolmogorov's
theorem for projective limits, an alternative construction of the measure.

As in (\ref{A.5}), for each $0<\alpha <1$ one defines a probability measure $%
\pi _{\mu ,\Lambda }^{\alpha }$ on $(\Gamma _{\Lambda },\mathcal{B}(\Gamma
_{\Lambda }))$ by%
\begin{equation}
\pi _{\mu ,\Lambda }^{\alpha }:=\sum_{n=0}^{\infty }\frac{E_{\alpha
}^{(n)}(-\mu (\Lambda ))}{n!}\mu _{\Lambda }^{(n)}  \label{B.5}
\end{equation}%
The family $\{\pi _{\mu ,\Lambda }^{\alpha }:\Lambda \in \mathcal{B}%
_{c}(M)\} $ of probability measures yields a probability measure on $(\Gamma
,\mathcal{B}(\Gamma ))$ with the $\pi _{\mu ,\Lambda }^{\alpha }$ as
projections, which being consistent uniquely determines a measure $\pi _{\mu
}^{\alpha }$ on $(\Gamma ,\mathcal{B}(\Gamma ))$ such that%
\begin{equation*}
\pi _{\mu ,\Lambda }^{\alpha }=\pi _{\mu }^{\alpha }\circ p_{\Lambda
}^{-1},\quad \forall \,\Lambda \in \mathcal{B}_{c}(M)
\end{equation*}

For the characteristic functional of the measure $\pi _{\mu }^{\alpha }$ one
obtains%
\begin{eqnarray*}
C_{\alpha }\left( \varphi \right) &=&\sum_{n=0}^{\infty }\frac{E_{\alpha
}^{\left( n\right) }\left( -\int_{\Lambda }d\mu \left( x\right) \right) }{n!}%
\left( \int_{\Lambda }e^{i\varphi \left( x\right) }d\mu \left( x\right)
\right) ^{n} \\
&=&\sum_{n=0}^{\infty }\frac{E_{\alpha }^{\left( n\right) }\left(
-\int_{\Lambda }d\mu \left( x\right) \right) }{n!}\int_{\Lambda
^{n}}e^{i\left( \varphi \left( x_{1}\right) +\varphi \left( x_{2}\right)
+\cdots +\varphi \left( x_{n}\right) \right) }d\mu ^{\otimes n} \\
&=&E_{\alpha }\left( \int_{\Lambda }(e^{i\varphi (x)}-1)\,d\mu (x)\right)
\end{eqnarray*}%
the last equality obtained by Taylor expansion of the Mittag-Leffler
function. Similarly to the $\alpha =1$ case, one sees that the probability
measure $\pi _{\mu }^{\alpha }$ on $(\mathcal{D}^{^{\prime }}(M),\mathcal{C}%
_{\sigma }(\mathcal{D}^{^{\prime }}(M)))$ is actually supported on
generalized functions of the form $\sum_{x\in \gamma }\delta _{x}$, $\gamma
\in \Gamma $.

One sees from (\ref{B.5}) that, instead of the uniform combinatorial weight $%
\frac{\exp (-\mu (\Lambda ))}{n!}$ for $n$ particles of the Poisson case ($%
\alpha =1$), one now has $\frac{E_{\alpha }^{(n)}(-\mu (\Lambda ))}{n!}$,
the rest being the same. Therefore the main difference in the fractional
case ($\alpha \neq 1$) is that a different weight is given to each $n$%
-particle space, although the support is the same. Different weights,
multiplying the $n$-particle space measures, may be physically significant
in that they have decays, for large volumes, smaller than the corresponding
exponential factor in the Poisson measure.

It is not surprising that the support of the measure $\pi _{\mu }^{\alpha }$
coincides with the support of the Poisson measure ($\alpha =1$). Using the
spectral representation of the Mittag-Leffler function%
\begin{equation*}
E_{\alpha }\left( -z\right) =\int_{0}^{\infty }e^{-\tau z}d\nu _{\alpha
}\left( \tau \right)
\end{equation*}
$\nu _{\alpha }$ being the probability measure in $\mathbb{R}_{0}^{+}$%
\begin{equation*}
d\nu _{\alpha }\left( \tau \right) =\alpha ^{-1}\tau ^{-1-1/\alpha
}f_{\alpha }\left( \tau ^{-1/\alpha }\right) d\tau
\end{equation*}%
and $f_{\alpha }$ the $\alpha -$stable probability density given by%
\begin{equation*}
\int_{0}^{\infty }e^{-t\alpha }f_{\alpha }\left( \tau \right) d\tau
=e^{-t^{\alpha }},\hspace{1cm}0<\alpha <1
\end{equation*}%
one may rewrite (\ref{B.4}) as%
\begin{equation*}
C_{\alpha }\left( \varphi \right) =\int_{0}^{\infty }\exp \left( \tau \int
(e^{i\varphi (x)}-1)\,d\mu (x)\right) \,d\nu _{\alpha }(\tau )
\end{equation*}%
the integrand being the characteristic function of the Poisson measure $\pi
_{\tau \mu }$, $\tau >0$. This shows that the characteristic functional (\ref%
{B.4}) coincides with the characteristic functional of the measure $%
\int_{0}^{\infty }\pi _{\tau \mu }\,d\nu _{\alpha }(\tau )$. By uniqueness,
this implies the integral decomposition%
\begin{equation}
\pi _{\mu }^{\alpha }=\int_{0}^{\infty }\pi _{\tau \mu }\,d\nu _{\alpha
}(\tau )  \label{B.6}
\end{equation}%
meaning that $\pi _{\mu }^{\alpha }$ is an integral (or mixture) of Poisson
measures $\pi _{\tau \mu }$, $\tau >0$.

A fractional Poisson analysis may be developed along the lines of the
inifinite-dimensional Poisson analysis \cite{Oliveira2}.

\textbf{Acknowledgment}

The author is grateful to Eric Carlen for an enlightening discussion on the
physical meaning of reducible functionals.

\end{document}